\documentclass[a4paper,11pt]{PoS}
\usepackage{subcaption}
\usepackage{amssymb,fontenc,times,mathptmx,amsmath,graphicx}

\smallskipamount

\title{Spontaneous Parity Violation}

\author{Goran Senjanovi\'c,$^{a,b}$ \\
\llap{$^a$}Arnold Sommerfeld Center, Ludwig-Maximilians University,
  Munich, Germany  \\
     \llap{$^b$}{International Centre for Theoretical Physics,
Trieste, Italy}
\\  E-mail:  \email{goran.senjanovic@physik.uni-muenchen.de}}

\author{Vladimir Tello,$^{b}$ \\
     \llap{$^b$}{International Centre for Theoretical Physics,
Trieste, Italy}
 \\ E-mail:  \email{vladimirtello@gmail.com}}

\abstract{We review here the central features of the idea of spontaneous breakdown of parity, in the context of its minimal realization, the Left-Right symmetric gauge theory. 
}

\FullConference{%
   Discrete 2022, 8th Symposium on Prospects in the Physics of Discrete Symmetries,\\
  7-11 November, 2022\\
  Baden Baden, Germany}

 \ShortTitle{Discrete 2022}

\begin{document}

\maketitle

\section{Introduction}
It is a familiar fact that the Standard Model (SM) is a remarkably successful theory of all  electro-weak interactions, charaterized by high precision. But what is the crux of it all? What is behind the SM's great simplicity and extraordinary predictivity?  The answer is twofold: 
\begin{itemize}
\item the Higgs mechanism, i.e. the  spontaneous symmetry breaking of the $SU(2)_L \times U(1)_Y$ gauge symmetry,  and
\item  the maximal violation of parity in weak interactions. 
\end{itemize}
The former implies that gauge bosons serve as the messengers of forces, and the Higgs mechanism guarantees the renormalizability of the theory. We all know this. However, the essential role of parity violation in the  functioning the Higgs mechanism, is often not emphasized enough. Let us recapitulate it now.

It all began with  Lee and Yang's question about parity non-conservation~\cite{Lee:1956qn} and the subsequent bombshell of Wu et al~\cite{Wu:1957my}: parity was maximally broken in beta decay. This discovery quickly lead to the V-A theory~\cite{Sudarshan:1958vf,Feynman:1958ty} according to which only LH fermions participate in the effective Fermi theory of weak interactions. This paved the way for Glashow's $SU(2)_L \times U(1)_Y$ gauge theory~\cite{Glashow:1961tr}, featuring an asymmetric fermion spectrum
 \begin{eqnarray}
q_L = \left( \begin{array}{c} u_L \\ d_L \end{array}\right), \,\,\,\,
\ell_L=
 \left( \begin{array}{c} \nu_L \\ e_L \end{array}\right)\,;\,\,\,\,\,\,\,\,\,\,\,\,\,\,\,\, u_R\,,\,\,d_R\,,\,\,e_R,
 \label{ds21}
\end{eqnarray}
In the words of  Weinberg, "V-A was the key"~\cite{Weinberg:2009zz} to understanding the weak interactions. The symmetry of the model implied massless gauge bosons and fermions, but in order to be renormalisable it needed the Higgs mechanism. Remarkably, a single Higgs doublet~\cite{Weinberg:1967tq} was sufficient to give masses to all particles: gauge bosons, charged fermions, and the Higgs itself. Through the magic of spontaneous symmetry breaking, masses now became dynamical parameters, related to physical processes. One particularly important consequence is that the masses of charged fermions determine  how the Higgs boson decays into fermion - antifermion pairs
\begin{equation} \label{hdecays}
\Gamma(h \to \bar f f) \propto \frac{m_f^2}{M_W^2} \, m_h\,,
\end{equation}
and similarly in the case of $W,Z$ gauge boson final states. Today, we have confirmed that the third-generations fermions and $W,Z$ gauge bosons owe their mass to the Higgs mechanism. With each new discovery and advancement,  we are getting closer to uncovering the origin of mass for all particles described by the Standard Model.

 
 However, despite its remarkable success the theory is incomplete in one importan aspect: it predicts massless neutrinos - the blessing of maximal parity violation that allows for the Higgs origin of charged fermion masses, is a curse when it comes to neutrinos.  Imagine that Lee and Yang had been wrong, and experiment had instead shown that parity was conserved. In such a left-right symmetric world the lepton assignment would take on a different form  
\begin{eqnarray}
\ell_L = \left( \begin{array}{c} \nu_L \\ e_L \end{array}\right), \,\,\,\,\,\,\,\,
\ell_R=
 \left( \begin{array}{c} \nu_R \\ e_R \end{array}\right).
 \label{ds21}
\end{eqnarray}
In this case, LR symmetry would imply the existence of $\nu_R$ and thus the neutrino would behave the same as the electron when it comes to mass  and would thus be massive. 

However, the Higgs mechanism would badly fail in such a case. First of all, the fermions would have direct mass terms, and the electron and neutrino would have the same mass. In order to split their masses, a real Higgs triplet would be needed, and by a fine-tuning the direct neutrino mass and the Yukawa induced one, it would be possible to keep neutrino light, but the $Z$-boson would remain massless. More Higgs scalars would be needed - say a doublet - but now the connection between Yukawa couplings and charged fermion masses would be lost completely. Moreover, there would be a problem of flavour violation in neutral currents, and, clearly,  Weinberg's classic~\cite{Weinberg:1967tq} would not have been written. 


In other words, maximal parity violation is necessary for the success of the SM in explaining the masses  of gauge boson and charged fermion, while parity conservation would automatically predict neutrino mass. How to reconcile these two dialectically opposite requirements? 
 The answer is almost obvious: assume that at the fundamental level, nature is actually left-right symmetric, and break parity  {\it  spontaneously}. 
 This idea, developed some fifty years ago, turned out to prophetically predict neutrino mass. 
 The question, though, was whether it was  a theory of the origin and nature of neutrino mass. By this we mean a true theory in the sense of Feynman - essentially a self-contained theory that makes unambiguous predictions. Its construction follows a logical structure which is very simple:

\begin{itemize}
\item Make a guess, for example, gauge principle.
\item Build a minimal formulation based on that guess.
\item Leave it at that so we can compute predictions.
\item Then let experiments test those predictions.
\end{itemize}

This approach sounds simple and obvious, yet very few models today satisfy these criteria. The Left-Right Symmetric model (LRSM) provides a minimal framework for spontaneous violation of parity and is, as we argue below, one such theory: a self-contained, complete theory of neutrino mass. It was suggested long before neutrinos were known to be massive. It is a theory of, not for, neutrino mass.

 \section{Spontaneous breakdown of Parity} 
  
 In the 1970's, a theory was proposed within the context of  the $SU(2)_L \times SU(2)_R \times U(1)_{B-L}$ gauge group augmented by the LR symmetry in  the form of parity~\cite{Pati:1974yy,Mohapatra:1974gc,Senjanovic:1975rk,Senjanovic:1978ev} (or alternatively charge conjugation). This program led to the prediction of a new set of gauge bosons, knows as the right-handed analogs of $W$ and $Z$ boson, which are now referred to as  the left-handed bosons.   The main features of this proposal can be summarised as:
     
     \begin{itemize} 
       \item Parity is predicted to be spontaneously broken~\cite{Senjanovic:1975rk,Senjanovic:1978ev}, with the scale of breaking corresponding to the mass of the $SU(2)_R$ gauge bosons $W_R$, with $M_{W_R} \gg M_{W_L} $ (the SM is recovered for $M_{W_R} \to \infty $).  
       Current experiments indicate that  $M_{W_R} \gtrsim 5 $TeV, and the predicted mass of the new neutral gauge boson $Z_R$ is even heavier:  $M_{Z_R} \simeq 1.7 M_{W_R}$. The discovery of $Z_R$  before the charged one would immediately disprove the theory.
      
  \item     In the process of the same SSB, the RH neutrinos also acquire large masses $m_N \propto   M_{W_R}$, which in turn give rise to small LH neutrinos through the seesaw mechanism~\cite{Minkowski:1977sc,Mohapatra:1979ia,Yanagida:1979as,GellMann:1980vs,Glashow:1979nm}. Due to the LR symmetry, the Dirac mass terms can be predicted~\cite{Nemevsek:2012iq,Senjanovic:2016vxw,Senjanovic:2018xtu,Senjanovic:2019moe} 
  from the light and heavy neutrino masses and mixings, which  enables the study of the orign of neutrino masses through the Higgs mechanism. This theory provides a plethora of calculable physical decays~\cite{Senjanovic:2018xtu} similar to the SM's predictions for charged fermion masses.
    
     \end{itemize}

 We refer the reader to~\cite{Senjanovic:2011zz,Tello:2012qda,Maiezza:2016ybz} for details on the Higgs sector and symmetry breaking in the LRSM. Here, we only provide a brief overview. Those unfamiliar with the field may benefit from a pedagogical explanation of the ideas discussed in ~\cite{Melfo:2021wry} . In the LRSM, quarks and leptons are completely symmetric under parity:
\begin{eqnarray}
q_L = \left( \begin{array}{c} u \\ d \end{array}\right)_L
& \stackrel{P}{\longleftrightarrow}&
 \left( \begin{array}{c} u \\ d \end{array}\right)_R = q_R\,,
 \nonumber \\
\ell_L = \left( \begin{array}{c} \nu \\ e \end{array}\right)_L
& \stackrel{P}{\longleftrightarrow}&
 \left( \begin{array}{c} \nu \\ e \end{array}\right)_R = \ell_R\,.
\label{dsLR21}
\end{eqnarray}

The relevant Higgs sector for SSB of Parity involves $B-L = 2$ triplets under $SU(2)_L$ ($\Delta_L$) and $SU(2)_R$ ($\Delta_R$)~\cite{Minkowski:1977sc,Mohapatra:1979ia}, with a potential written schematically (ignoring the triplet structure)
      \begin{equation}
   {\cal V}_\Delta = -\mu^2 ( \Delta_L^2 + \Delta_R^2) + \lambda (\Delta_L^2 + \Delta_R^2)^2 + \lambda ' \Delta_L^2 \Delta_R^2  .
   \label{delta-lag-1}
   \end{equation}
It follows that
     \begin{equation}
 \lambda ' \geq  0 \Rightarrow \langle \Delta_L \rangle = 0 \, , \; \langle \Delta_R \rangle \neq 0,
   \label{delta-lag-1}
   \end{equation}
   leading to the  spontaneously breakdown of parity. The unbroken symmetry defines left-handedness and this schematic analysis can be  extended to a rigorous analysis with the full $SU(2)$ quantum numbers.
   
   To account for the SM symmetry breaking, the Higgs sector is complemented by a bi-doublet (doublet under both $SU(2)_L$  and $SU(2)_R$), denoted by $\Phi$, containing the usual Higgs doublet. As a result, the Yukawa interactions take the form
 \begin{equation}
   {\cal L}_Y = Y_\Phi \bar f_L \Phi f_R + Y_{\tilde \Phi} \bar f_L  \tilde\Phi f_R  + Y_\Delta ( \ell^T_L \, \sigma_2 \Delta_L C \, \ell_ L + \ell^T_R \, \sigma_2 \Delta_R C \, \ell_ R),
   \label{yukawa-LR}
   \end{equation}
  where $\tilde\phi = \sigma_2 \Phi^* \sigma_2$.   
  
     The theory delivered at once - at the time when neutrino was believed massless - a fundamental prediction: neutrinos had to be massive just as their charged cousins.  In addition to the Dirac mass triggered by the vev of $\phi$ (which give mass to charged leptons and quarks)
     \begin{equation}
     M_D = Y_D \langle \phi\rangle \, ,
     \end{equation}
     there is also a Majorana mass term for RH neutrinos which arises from the RH triplet vev
       \begin{equation}
     M_N = Y_\Delta \langle \Delta_R \rangle \, ,
     \end{equation}
          In the above, $Y_D$ and the SM doublet $\phi$ are linear combinations of $Y_{\Phi,\tilde\Phi}$ and the $SU(2)_L$ doublets in $\Phi$, respectively
          
          The LRSM predicted the existence of massive neutrinos at a time when they were believed to be massless, but it took about a quarter of a century before neutrino mass was established experimentally. 
           In this version of the LRSM, via the see-saw mechanism, the smallness of neutrino mass is directly tied to near-maximality of parity violation. This means that the SM becomes simply a limit when the right-handed (RH) gauge boson interactions decouple in the large mass regime.  The Majorana masses for neutrinos imply the existence of one heavy neutrino state $N$ ($\sim \nu_R$), and a light one, $\nu$ ($\sim \nu_L$),  with
     \begin{equation}
     M_\nu   \simeq   - M_D^T  \frac{1}{M_N} M_D,  \quad   M_N  \propto  M_{W_R}\,. 
     \label{seesaw}
     \end{equation}

A crucial question is, of course, the scale of the theory. For all we know, since neutrino mass is so tiny, the SM
could be valid all the way up to astronomically high energies on the order of $10^{14}$ GeV, where the seesaw scale or the cutoff scale of the d=5 effective operator of neutrino mass could set in. However, this does not necessarily mean that a change is not just around the corner, and that it may be accessible even at the LHC. 

The possible discovery of a deviation in the $W$-mass from the Standard Model value by the CDF collaboration would be a strong indication of new physics at the TeV scale, and could point to the existence of right-handed gauge bosons and heavy neutrinos.  Time will tell whether this is the case, but in the meantime, it is worth investigating the possibility of reaching the scale $M_R$.

     The bottom line is that neutrino mass, by itself, makes  a strong case for the spontaneous breakdown of parity. But there is more to it:
    in  recent years, it has been shown that the LRSM is indeed a self-contained predictive theory of neutrino mass, analogous to the SM as the theory of charged fermion mass.  The Majorana nature of neutrino plays a crucial role in probing the theory, as it implies Lepton Number  Violation (LNV)  and allows for two important processes that are  actively being pursued: neutrinoless double-beta decay and LNV at colliders. The crucial point is that if neutrinoless double-beta decay is observed, it may actually indicate that the LR scale could be accessible at near-future colliders, if not at the LHC itself.

 \section{Lepton Number Violation at low and high energies}

\begin{figure}[h]
\centerline{
\includegraphics[width=.32\columnwidth]{
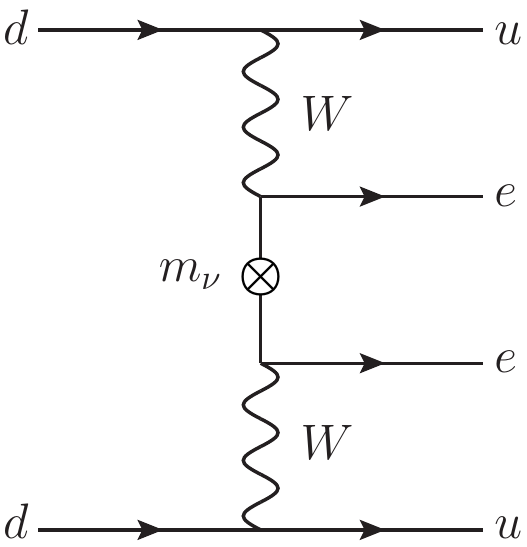}  }
\caption{Neutrinoless double-beta decay through the exchange of Majorana neutrino.}
\label{fig-0nu2beta}
\end{figure}

\begin{figure}[h] 
   \centering
    \includegraphics[width=0.47\textwidth]{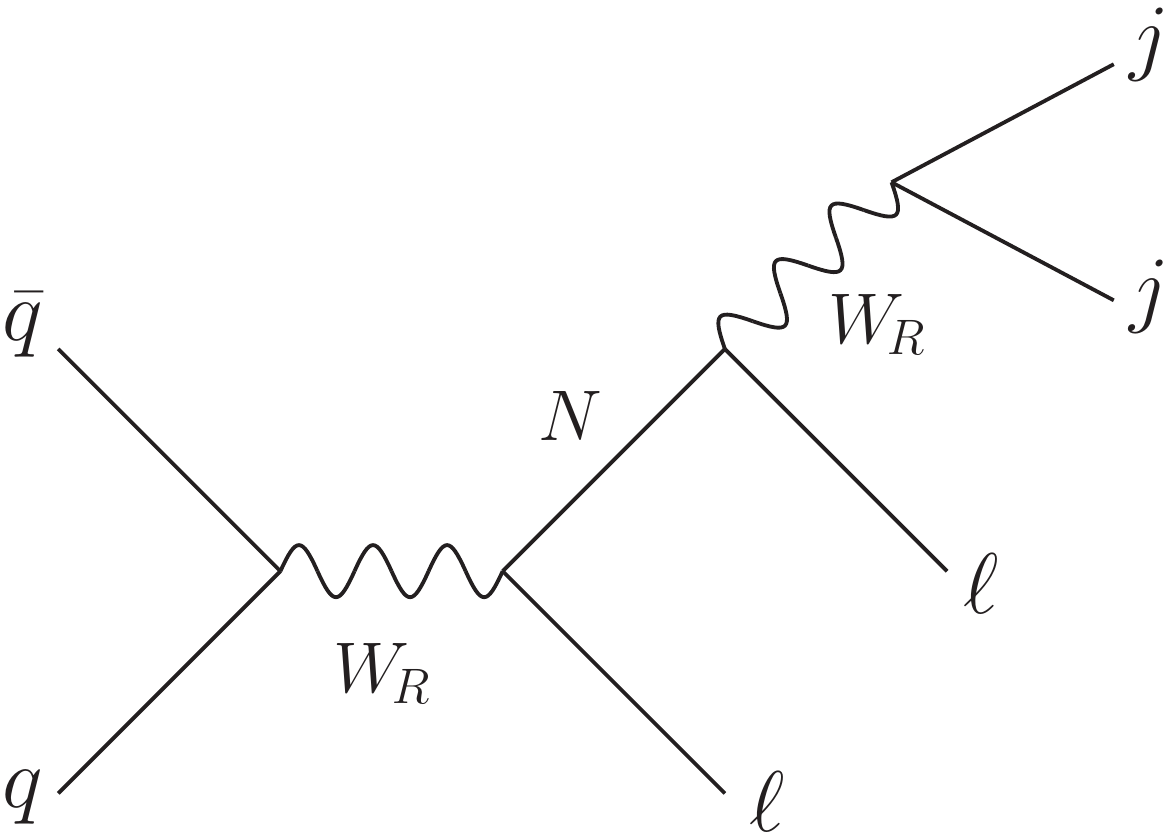}
  \caption{The Keung-Senjanovi\'c process at hadron colliders}
  \label{fig-ks}
\end{figure}

A Majorana mass for the neutrino would allow for neutrinoless double-beta decay, as shown in Fig.\ref{fig-0nu2beta}.
The signature is two left-handed electrons $e_L$. For $p\simeq 100$MeV, the amplitude is given by
\begin{equation}
{\cal A}_\nu \propto \frac{G_F^2 m_\nu^{ee}}{p^2} \simeq G_F^2 10^{-8} \text{GeV}^{-1}.
\end{equation}
The current limits~\cite{GERDA:2020xhi} for this process, $ \tau_{0\nu2\beta} \gtrsim 10^{26} yr$, implies a limit on the light neutrino Majorana mass
\begin{equation}
m_\nu^M \lesssim 0.3 \text{eV}.
\end{equation}
What if some new physics were behind this process? Let us consider a general effective $d=9$ operator, suppressed by a  scale $\Lambda$
\begin{equation}
\frac{1}{\Lambda^5} n\,n\,\bar p\,\bar p\,\bar e\,\bar e, 
\end{equation}
where $n$ and $p$ are the neutron and proton fields, respectively. The current limits on $0\nu2\beta$ then give $\Lambda \geq 3 $TeV, which is accessible at LHC energies. In LRSM, the heavy neutrino could do the job~\cite{Mohapatra:1979ia,Mohapatra:1980yp}, and the outcome would be two right-handed leptons. This would be related to the lepton number violation at hadron colliders, and the 
direct probe of the Majorana nature of the RH neutrino is also possible, through the Keung-Senjanovi\'c (KS) process~\cite{Keung:1983uu} shown in Fig. \ref{fig-ks}.

%

The KS process has two important aspects. The first aspect is the direct LNV at hadron colliders: in complete analogy with the (LH) $W$ production, once $W_R$ is produced it decays into a lepton and a RH neutrino $N$. The Majorana nature of $N$ can then lead to LNV, and you get same-sign leptons as in $0\nu2\beta$ above (this is a generic feature of models of neutrino Majorana mass, see~\cite{Senjanovic:2011zz}).  This process being a collider process, it allows for the  reconstruction of the properties of $W_R$ and $N$ from the information on the final states.  
The second aspect is that, since $N$ is Majorana, being half particle and half anti-particle, once you produce it on-shell it has to decay equally often into a charged lepton and its anti-particle~\cite{Keung:1983uu}. Therefore, the outcome of the process is 50\% same-sign leptons and 50\% lepton-antilepton.

The LRSM can provide a deep connection between $0\nu2\beta$  and the LHC~\cite{Tello:2010am}.
Additionally, there is also a remarkable connection with low energy lepton flavor violation processes, such as $\mu \to e \gamma$, $\mu \to e \,e \, \bar e$ and $\mu \to e $ conversion in nuclei.  It has been shown that these processes are correlated in the LRSM~\cite{Cirigliano:2004mv}. For the possibly accessible LR symmetric scale in the TeV region, their suppression implies $m_N \ll M_{W_R}$ (or an unnatural degeneracy of $N$ masses), in analogy with the small charm quark mass as a guarantee for tiny $K - \bar K$ mass difference. This has been summarised in~\cite{Tello:2012qda}.

Once the Majorana nature of $N$ is verified,  one can turn to probing the nature of the LH neutrino mass by disentangling the see-saw, as we now discuss.

\section{Untangling the see-saw}

A crucial aspect of the LRSM is its self-containment. To begin with, the seesaw can be untangled and $M_D$ determined from \eqref{seesaw}. In recent years, a lot of effort has been devoted to this question~\cite{Senjanovic:2018xtu}. Once the masses  and mixings of RH and LH neutrinos are measured, the neutrino Dirac mass can be determined. 
This is best illustrated for the case of charge conjugation instead of parity, which then gives
 \begin{equation}\label{untangle}
 M_D = i M_N \sqrt  {M_N^{-1} M_\nu}.
 \end{equation}
All the ambiguity that normally plagues~\cite{Casas:2001sr} the seesaw is gone.To see this, compare this with the naive seesaw result, where
 \begin{equation}\label{naive}
 M_D = \sqrt{m_N}  O \sqrt{M_\nu},
  \end{equation}
is expressed in terms of a completely arbitrary complex orthogonal matrix $O$. In other words, once the KS process  allows to determine $M_N$,  the matrix $M_D$ can be determined without any ambiguity from \eqref{untangle}, by the reason of that $M_\nu$ is being probed (we already know all three mixing angles) in low energy processes. The same result has been found in the case where of parity itself, albeit with a greater deal of calculational tedium.

This, moreover, allows for precise predictions in a  plethora of new decays~\cite{Senjanovic:2018xtu}, all depending on $M_D$ and/or $M_N$. In particular, one can predict the decay of RH neutrino into a $W$ and a charged lepton~\cite{Nemevsek:2012iq}, illustrated here for a simple case of the same LH and RH leptonic mixings
\begin{equation} \label{Ndecays}
\Gamma(N \to W \ell) \propto \frac{m_N^2}{M_W^2} \, m_\nu\,.
\end{equation}
For an associated LHC study, see e.g.~\cite{Arbelaez:2017zqq}. Since this decay is induced by the $\nu - N$ mixing, the outgoing charged lepton must be left-handed. However, there is also an associated decay into the right-handed charged lepton due to $W_L - W_R$ mixing, and so it is important to be able to measure the lepton chirality. This was studied in~\cite{Ferrari:2000sp}, and argued to be feasible at the LHC.

Notice that this is in complete analogy with the SM, where the knowledge of charged fermion masses gives you the Higgs decay rates
\begin{equation} \label{hdecays}
\Gamma(h \to \bar f f) \propto \frac{m_f^2}{M_W^2} \, m_h\,.
\end{equation}
Arguably, the LRSM does for neutrinos what the SM did for charged fermions: it allows us to probe the Higgs origin of mass.

\section{Quark Sector}

\begin{figure}[h] 
   \centering
  \begin{subfigure}[b]{0.45\textwidth}
    \includegraphics[width=\textwidth]{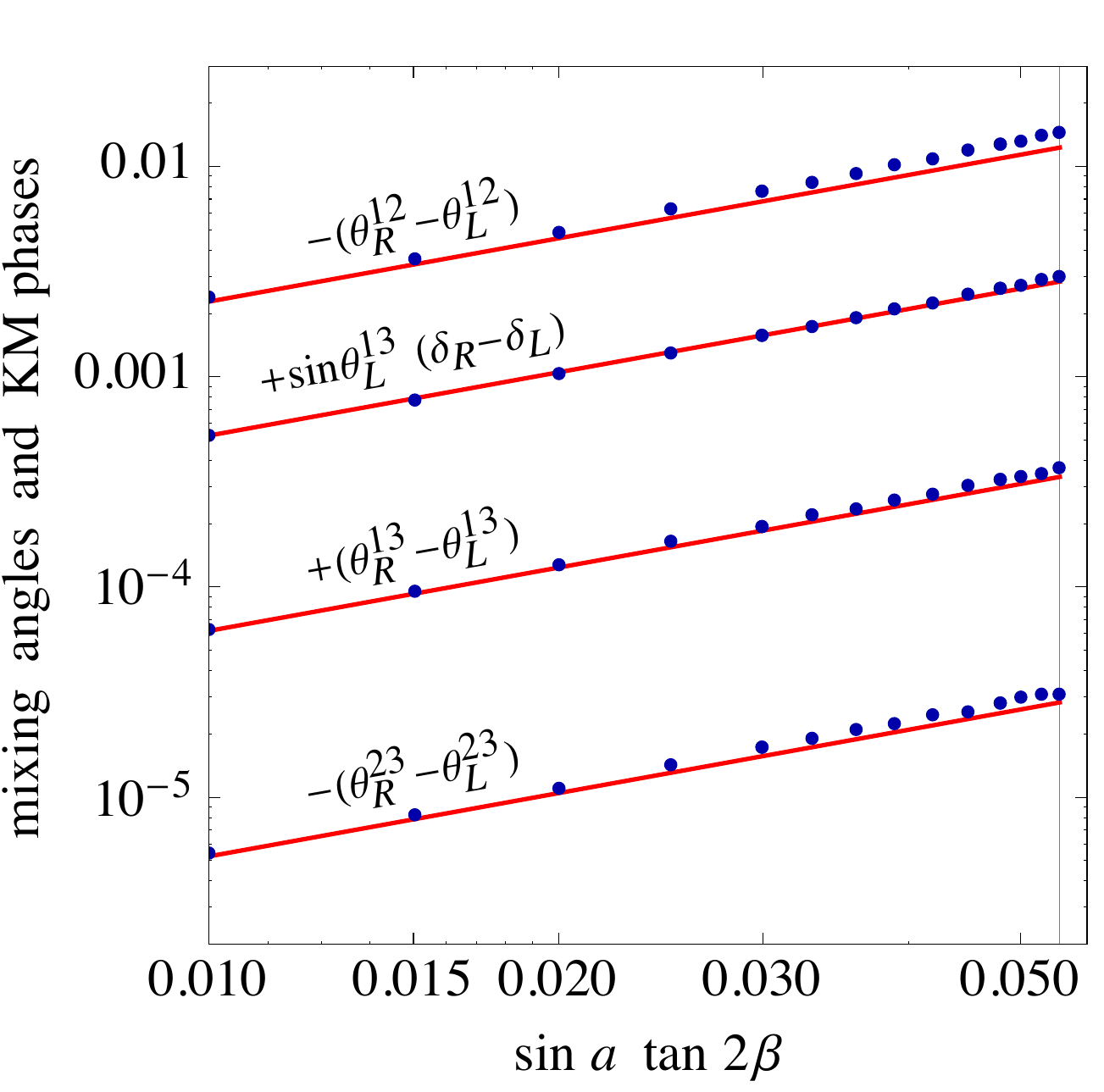}
    \caption{Right and left mixing angles differences.}
    \label{fig:VR-angles}
    \end{subfigure}  
  \begin{subfigure}[b]{0.45\textwidth}
    \includegraphics[width=\textwidth]{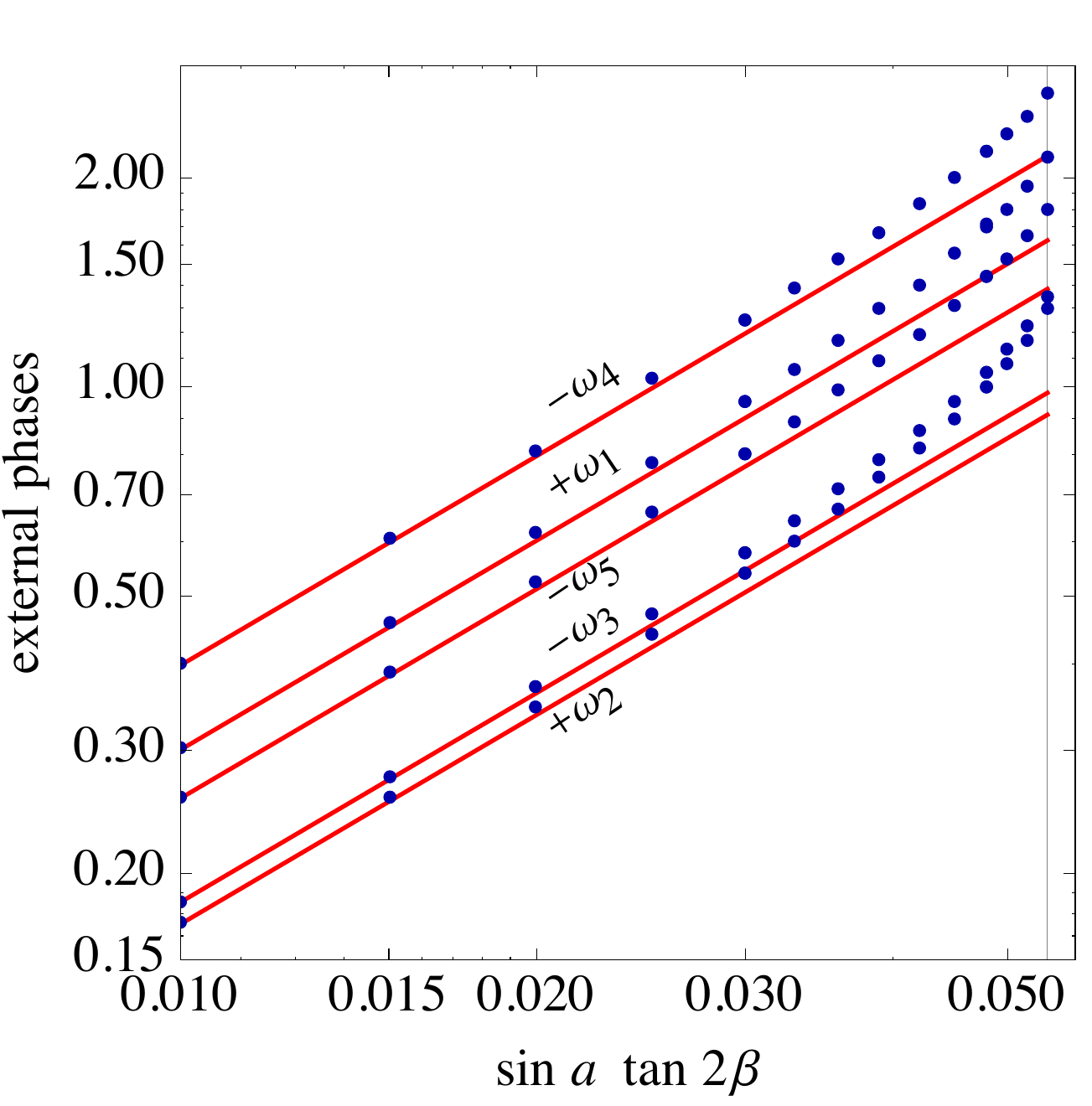}
    \caption{Right handed external phases.}
    \label{fig:VR-phases}
      \end{subfigure}  
  \caption{Parameters of $V_R$ as a function of $\epsilon=$
 sin a tan 2 $\beta$.}
  \label{fig:VR-angles-phases}
\end{figure}

The determination of the right-handed quark mixing is a crucial aspect of the Left-Right Symmetric Model, as it allows for precise calculations in processes involving the interactions of right-handed gauge bosons with quarks. The analytic expression for the right-handed quark mixing matrix had been a challenging task for almost four decades since the inception of the theory. However, recent developments have led to its exact equation and solution in terms of a series expansion of a small parameter~\cite{Senjanovic:2014pva,Senjanovic:2015yea}. The mixing matrix $V_R$ turns out to be a function of the CKM matrix, quark masses, and a small CP-violating parameter only
\begin{equation} \label{eq:master}
(V_R)_{ij} \simeq (V_L)_{ij} - i \epsilon \frac{(V_L)_{ik} (V_L^\dagger m_u V_L)_{kj}} {m_{d_k} + m_{d_j}}.
\end{equation}
 Remarkably, the predicted near equality of left and right-handed mixing angles, despite maximal breaking of parity in weak interactions, is a characteristic feature of the theory that can be probed by the LHC.  This can be seen through a straighforward calculation using the parametrization given by
 \begin{equation} \label{pdg-parametrization}   
\!\!V_R\! \equiv\! \text{diag}(e^{i\omega_1}\!,e^{i\omega_2}\!,e^{i\omega_3}\!)\,V(\theta_{ij}^R,\delta_R)\, \text{diag}(e^{i\omega_4}\!,e^{i\omega_5}\!,1),
\end{equation}
which generates the following expression for the differences between left and right-handed mixing angles:
 \begin{align}\label{eq:difference12}
   \theta^{12}_R - \theta^{12}_L &\simeq  - \epsilon \frac {m_t}{m_s} s_{23} s_{13}   s_\delta,
\\[3pt]
   \label{eq:difference23}
   \begin{split}
\theta^{23}_R - \theta^{23}_L &\simeq -  \epsilon  \frac{m_t}{m_b}\frac{m_s}{m_b} s_{12} s_{13} s_\delta,
\end{split}
\\[3pt]
    \label{eq:difference13}
    \begin{split}
     \theta^{13}_R - \theta^{13}_L &\simeq -\epsilon \frac{m_t}{m_b} \frac{m_s}{m_b} s_{12} s_{23} s_\delta,\end{split}
 \end{align}
 %
 %
 where $s_{ij}=\sin \theta^L_{ij}$, $c_{ij}=\cos \theta^L_{ij}$.  In Fig.~\ref{fig:VR-angles}  we have plotted in red lines these first order results, and with blue dots the numerical solutions of the exact equation.   The first order is an excellent approximation and the agreement between the two is manifest through the  whole physical range of $\epsilon$.
 In other words, for $\epsilon \ll 1$, one obtains  
\begin{equation}
\theta_R \simeq \theta_L.
\end{equation} 
 This important result holds true fo the same reason that makes the SM work so well when it comes to meson mixings and CP violation: small CKM mixing angles. An immediate consequence of this is that the quoted limits on $M_{W_R}$, which assume the same left and right-handed mixings, are now well- justified. 
  
Furthermore, there are strong correlations between  the phases in the RH sector and the CKM phase, allowing to make predictions for CP violation beyond the SM, if $W_R$ was to be found at the LHC. 
Specifically,
 %
%
 \begin{align}\label{eq:extphasesw1} 
\omega_1  &    \simeq     \epsilon\bigg(\frac{m_t}{2m_b}+  \frac{m_c c_{23}^2 + m_ts^2_{23}}{m_s} 
     - \frac{m_dc_{12}^2 + m_s s^2_{12}}{2m_u}  \bigg) ,
   \\[5pt] 
    \label{eq:extphasesw2} 
  \omega_2&  \simeq \epsilon \frac{m_t}{2m_b},\quad    \omega_3   \simeq -  \epsilon \frac{m_t}{2m_b},
        \\[5pt]
     \label{eq:extphasesw4} 
    \omega_4   & \simeq  -  \epsilon\bigg( \frac{m_t}{2m_b}+ c_{12}^2\frac{m_c c_{23}^2 + m_ts^2_{23}}{m_s} + s^2_{12}\frac{m_c c_{23}^2  + m_ts_{23}^2}{2m_d} \bigg) ,
   & \\[0pt]
    \label{eq:extphasesw5} 
     \omega_5 & \simeq  - \epsilon \left(\frac{m_t}{2m_b}+ \, c_{12}^2\frac{m_c c_{23}^2 +m_ts^2_{23}}{2m_s}  \right),
  \end{align}
 and similarly for the KM phases
 \begin{align} 
  \label{eq:Delta}
  \delta_R - \delta_L  &\simeq  \epsilon \frac{m_cc_{23}^2 +m_t s_{23}^2}{m_s}.
  \end{align}
We have plotted these phases in Fig.~\ref{fig:VR-phases}. Again, the first-order results are shown in red, while the numerical results are shown in blue. Notice that these results start to deviate for larger values  $\epsilon \gtrsim 0.03 $, which simply implies the need for higher order terms in \eqref{eq:master}.

Bottom line, the RH quark mixing mass matrix is completely determined by the CKM LH one. This is a profound result: in spite of near maximal violation of parity in low energy weak interaction, the memory of the original symmetry remains almost intact when it comes to quark mixing matrices. The reason is parity being spontaneously broken, but not only - an important role is played by the fact that the LH. mixing angles are small, especially with the thrid generation.
Thus th study of the right-handed quark mixing provides a fundamental aspect of the Left-Right Symmetric Model, and its precise determination opens up new avenues for testing the symmetries and predictions of this theory.

\section{LHC searches}

The search for the right-handed $W_R$ gauge boson is one of the objectives of the LHC experiments and its discovery would provide a boost for  the LRSM and the idea of spontaneous breakdown of parity.
Although this search falls under the generic search for $W'$, there is more to it here. One of the central features of the LRSM, as we have seen,  is that the couplings of $W_R$ with quarks are related to the CKM mixings. This simplifies the search for $W_R$ at the LHC, as it allows for the determination of the production cross section and decay modes of $W_R$ as a function of the CKM mixings. While this assupmtion is often made   {\it ad-hoc}   for any $W'$,   it would be a kind of miracle in general, as it requires a memory of the original LR symmetry associated with the spontaneous breaking.

The current experimental limits on the mass of $W_R$ come from direct searches at the LHC. The CMS and ATLAS collaborations have performed searches for $W_R$ in the dijet channel, where $W_R$ is produced via quark-antiquark annihilation and decays into two jets. The most recent results set a lower limit of $M_{W_R} \geq 4.4$TeV \cite{ATLAS:2021drn, CMS:2023pas}. These limits are based on the analysis of the dijet invariant mass distribution and are valid for a $W_R$ that decays exclusively to quarks. It is worth noting that they  depend crucially on  the left and right-handed quark mixings being the same, and may change significantly if this is not the case.
In short, the search for the right-handed $W_R$ gauge boson is an active area of research at the LHC, and the current limits on its mass provide important constraints on the left-right symmetry breaking. Further experimental and theoretical studies are needed to fully explore the implications of the LRSM and to shed light on the origin of the electroweak symmetry breaking.

At the same time, the search for right-handed neutrinos - as the source of neutrino mass through the seesaw mechanism - at the LHC has been a topic of great interest in recent years. The golden channel for detecting these particles, as discussed above, is the  KS process~\cite{Keung:1983uu}.  This process leads to a final state with two charged leptons and two jets, tailor-made for the LHC, and pursued both by the ATLAS and CMS experiments. These searches have placed constraints on the masses of the right-handed neutrinos and the $W_R$ gauge boson.

\begin{figure}[h] 
   \centering
  \begin{subfigure}[b]{0.45\textwidth}
    \includegraphics[width=\textwidth]{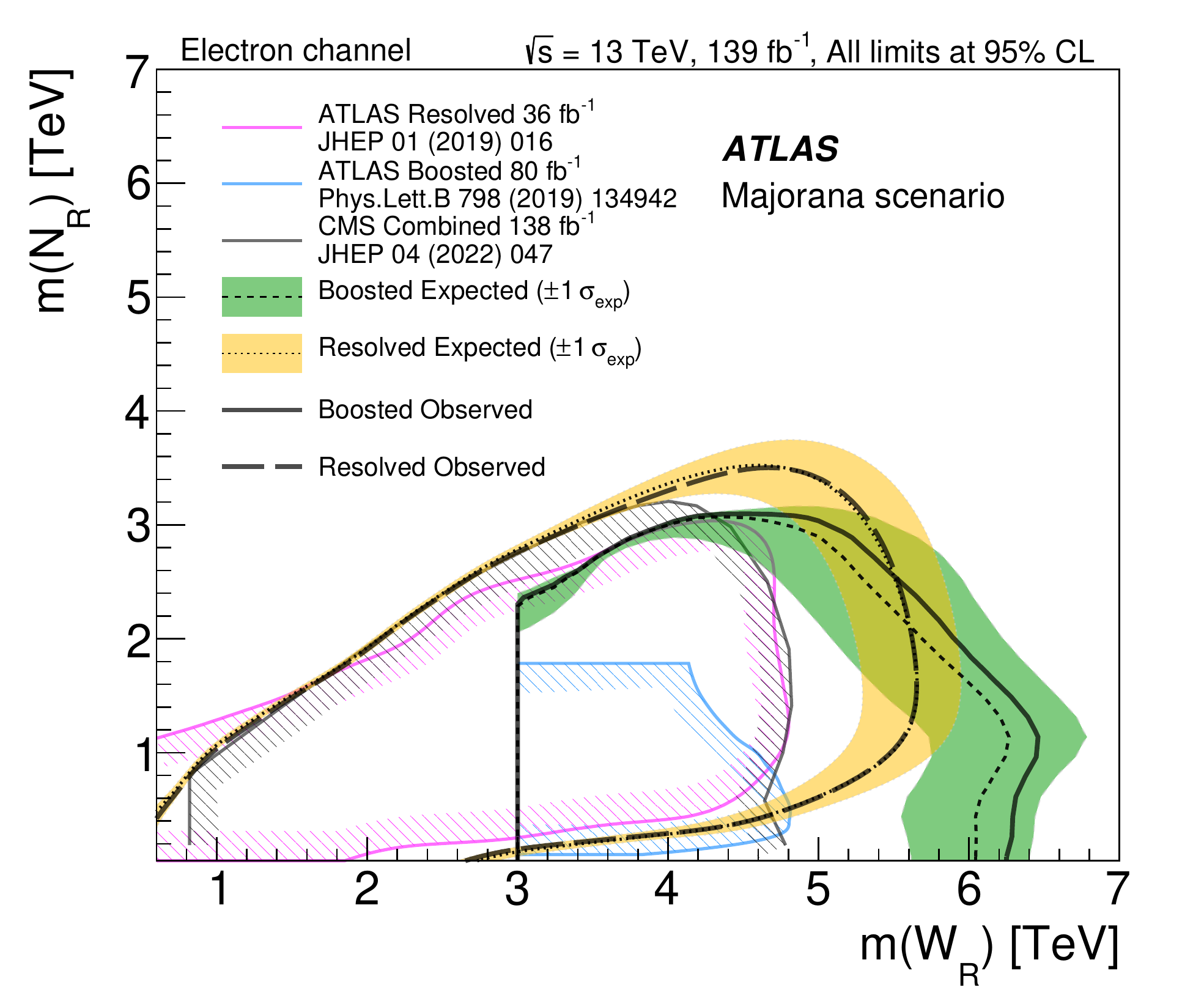}
    \caption{Atlas reach, ee channel. Taken from \cite{ATLAS:2023cjo}.}
    \label{fig:KS-Atlas}
    \end{subfigure}  
  \begin{subfigure}[b]{0.5\textwidth}
    \includegraphics[width=\textwidth]{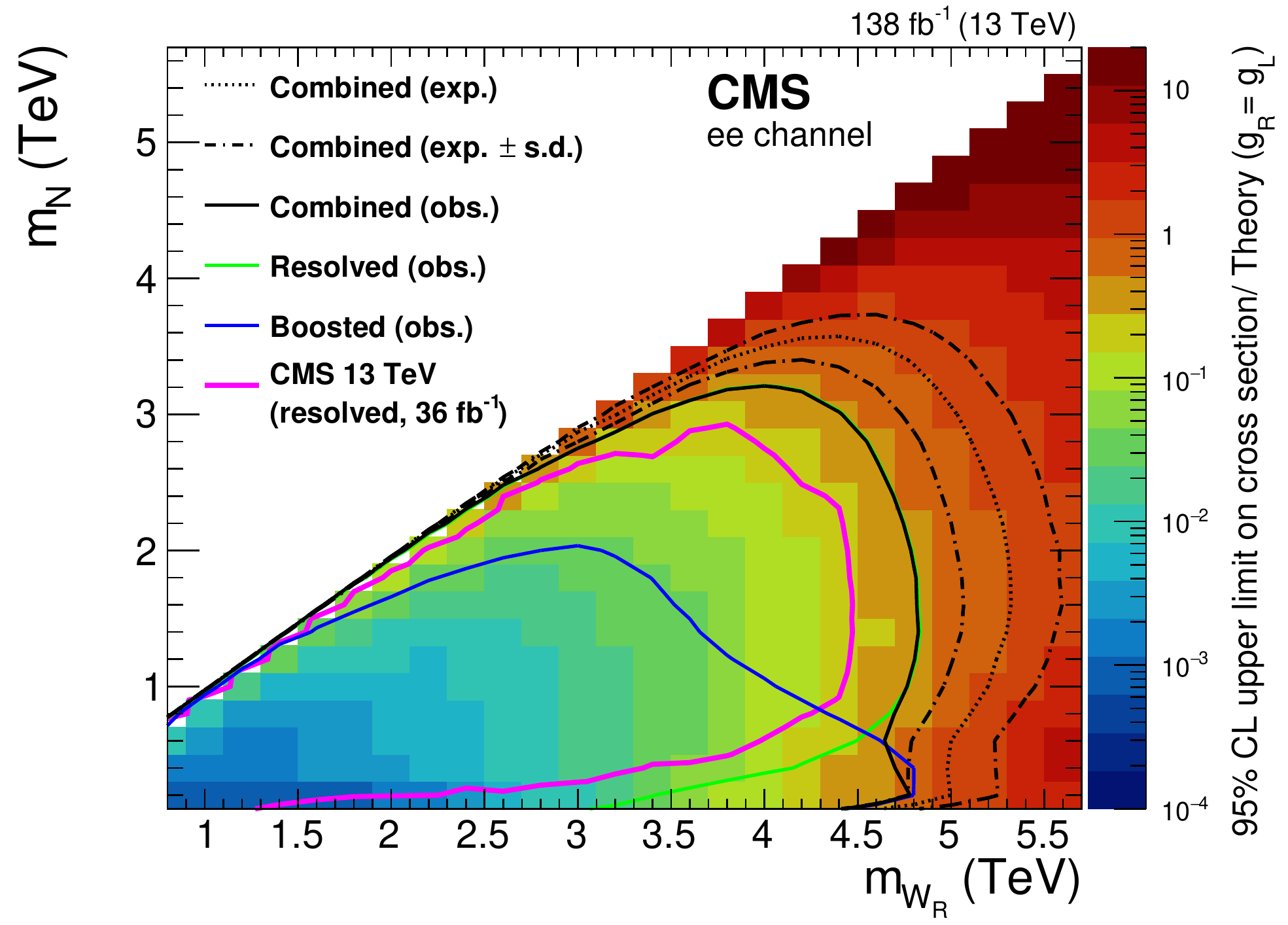}
    \caption{CMS reach, ee channel. Taken from 
   \cite{CMS:2021dzb}.}
    \label{fig:CMS-Atlas}
      \end{subfigure}  
  \caption{KS searches at the LHC.}
  \label{fig:KS-LHC}
\end{figure}

 Figure \ref{fig:KS-LHC} provides the most recent  results of the ATLAS detector reach for $W_R$ and right-handed neutrino masses, highlighting the potential of the golden channel proposed by Keung and Senjanovic for probing the right-handed neutrino masses up to 3.8 TeV. It can be seen readily that the LHC can discover the $W_R$ with the mass up to roughly 6 TeV. Besides direct lepton number non-conservation and the determination of the RH neutrino masses and mixings, this could have profound consequence for the neutrinoless double decay, which could then be mediated by $W_R$ through the RH neutrino Majorana mass.

 The search for right-handed neutrinos in the KS golden channel remains an active area of research at the LHC. With the ongoing data taking and future upgrades to the LHC, it is hoped that even stronger limits will be placed on the masses of these particles and their couplings to the SM particles. A comprehensive analysis of the LHC potential for the $W_R$ search, with the emphasis on the KS process, was performed in~\cite{Nemevsek:2018bbt}.


\section{Summary and Outlook}
The idea of parity restoration at a fundamental level is not new. It was there in the Lee and Yang classic that suggested parity violation in the first place. They imagined a possibility of what is called today mirror fermions, ruled out in the context of the SM. About fifty years ago,   at the very beginning of the theoretical and phenomenological success of the SM, this idea was taken seriously in the context of a left-right symmetric gauge model with an immediate consequence of non-vanishing neutrino mass. Eventually, with the advent of the seesaw mechanism, it turned out that the smallness of neutrino mass was directly associated with the near maximality of parity violation in the weak interactions. In the SM limit, as the LR scale going to infinity, neutrino mass go to zero as it has to be. The LRSM was a prophetic theory of neutrino mass, and    recent years  have shown that it is a completely self-contained theory of the origin and nature of neutrino mass. Just as the SM allows for the Higgs probe of charged fermion masses through associated Higgs boson decays, the LRSM does basically the same for neutrino, albeit with more complexity.

The essence in all of this is the Majorana nature of the  heavy RH neutrino, which through the seesaw gives rise to a Majorana mass for the LH neutrino. The Majorana nature of RH neutrinos can be  directly tested at hadronic colliders through the Keung-Senjanovic process, since their decays must be both lepton number violating and conserving, in equal proportions. This allows for the determination of their masses and mixings, just as the low energy processes do for the LH neutrinos. In turn, the original LR symmetry of the theory, be it parity or charge conjugation, then uniquely determines the Dirac mass terms (or equivalently Yukawas) which in turn amounts to predicting a plethora of particle decays, both of RH neutrinos and associated Higgs bosons. The LRSM, surprisingly in a sense, is a truly predictive theory in a sense of Feynman, as argued in the Introduction. If the reader has gone through this short, almost telegraphic, review of the central aspects of the theory and its phenomenological consequences, they ought to be convinced of our claims. If not, we encourage them to go through the original papers  cited here, that led to these claims. 

We conclude with two important questions related to the LRSM theory: could   the LRSM provide the sources of dark matter and leptogenesis,    and if yes, what are the constraints on the theory's parameter space?  The answer the first question is yes, with important constraints on the parameter space of the  theory. The lightest RH neutrino could  in principle provide a sufficiently long lived dark matter candidate~\cite{Bezrukov:2009th,Nemevsek:2012cd}, if  its mass was roughly keV~\cite{Dodelson:1993je}. This would, of course,  rule out the KS process, but it should still be kept in mind as a possibility. It moreover requires a $W_R$ mass   beyond the LHC reach (except for the tiny window around 5 TeV  which is about to be closed). On the other hand, if leptogenesis were the source of the universe's baryon excess, an even stronger limit on the order of 30 TeV~\cite{Frere:2008ct} would emerge.
Thus, if $W_R$ were to be discovered at the LHC, one would have to look elsewhere for the explanation of baryogenesis and 
 the KS process would lose its beauty and its possible deep connection with the neutrinoless double beta decay.

{\bf Acknowledgements} G.S. wishes to thank the organisers of the Discrete 22 Conference for an invitation to present ideas discussed here. We are deeply grateful to Alejandra Melfo for her help with the manuscript and to Damir Lelas and Fabrizio Nesti for helping us with the relevant experimental limits. 


\bibliographystyle{JHEP}
\bibliography{biblio.bib}

\end{document}